\documentclass[prb,aps,prb,amsmath,floatfix,twocolumn,showkeys]{revtex4}
\usepackage[]{graphicx}
\usepackage{graphicx}
\usepackage{dcolumn}
\usepackage{bm}

\begin{document}
\title{Transient thermoelectricity in a
vibrating quantum dot in Kondo regime}
\author{A. Goker$^{1}$ and B. Uyanik$^{2}$}

\affiliation{$^1$
Department of Physics, \\
Bilecik University, \\
11210, G$\ddot{u}$l$\ddot{u}$mbe, Bilecik, Turkey
}

\affiliation{$^2$
Department of Physics, \\
Cukurova University, \\
01330, Balcali, Adana, Turkey
}

\date{\today}

\begin{abstract}
We investigate the time evolution of the thermopower
in a vibrating quantum dot suddenly shifted into the 
Kondo regime via a gate voltage by adopting the
time-dependent non-crossing approximation and linear 
response Onsager relations. Behaviour of the instantaneous
thermopower is studied for a range of temperatures both 
in zero and strong electron-phonon coupling. We argue that
inverse of the saturation value of decay time of thermopower 
to its steady state value might be an alternative tool in
determination of the Kondo temperature and the value of the
electron-phonon coupling strength.    
\end{abstract}



\thispagestyle{headings}

\maketitle
Investigation of time-dependent electron transport 
in single electron transistors has gained considerable 
traction lately as a result of tremendous advances
in the burgeoning field of nanotechnology and their
perceived potential to replace MOSFET transistors
\cite{semiconductor} one day. Development of quantum 
computers \cite{ElzermanetAl04Nature} and single electron 
guns \cite{FeveetAl07Science} are expected to benefit 
from advances in detection of electrons in real time 
too \cite{LuetAl03Nature}.

Transient current ensuing after sudden shifting 
of the gate or bias voltage \cite{NordlanderetAl99PRL,
PlihaletAl00PRB,MerinoMarston04PRB} displays different
time scales \cite{PlihaletAl05PRB,IzmaylovetAl06JPCM}. 
For an asymmetrically coupled system, interference
between the Kondo resonance and the sharp features
in the contacts' density of states gives
rise to oscillations in the long time scale 
\cite{GokeretAl07JPCM}. Modeling the contacts' 
density of states in a realistic fashion via 
ab initio calculations yielded accurate 
predictions about transient current
\cite{GokeretAl10PRB,GokeretAl11CPL}.

Measurement of thermopower (Seebeck coefficient) $S$
can provide additional insight into transport experiments
because its sign is a valuable tool to determine the 
alignment of orbitals of the impurity with respect
to the Fermi levels of the contacts. To this end,
significant progress has been achieved in performing
thermoelectric measurements in molecular junctions
\cite{Reddyetal07Science,Bahetietal08NL,Malenetal09NL,TangetAl10APL}.

Theoretical studies focused on incorporating
strong correlation effects into this prototype.
Kondo effect, arising from a hybridization between
the net spin localized within the impurity and the 
electrons in the contacts, is a prime example
of strong electronic correlations. Schemes employing
Ng's ansatz\cite{DongetAl02JPCM} and Wilson's
numerical renormalization group\cite{CostietAl10PRB} 
both found that the sign of the thermopower can be
adjusted by tuning the energy level of the quantum 
dot in the presence of Kondo correlations. When the
electrodes are ferromagnetic, it was found that the 
thermopower is suppressed at low temperatures in
parallel configuration and asymmetrically coupled
antiparallel configuration \cite{KrawiecetAl06PRB}.

Nevertheless, all of the aforementioned studies
only took into account the steady state behaviour 
of thermoelectric transport and there was very 
little understanding of the temporal evolution
of thermopower until now. Indeed, a recent work
made the first attempt to elucidate the time-dependent 
behaviour of the Seebeck coefficient for a noninteracting
system \cite{CrepieuxetAl11PRB}. In this paper, we 
will take a step forward and study the transient 
response of thermopower for an interacting quantum 
dot suddenly moved into the Kondo regime with a 
time dependent gate voltage.

We can describe this device with Holstein
Hamiltonian. Its three pieces representing
the contacts, the quantum dot and the tunneling
between them can be written respectively as
\begin{equation}
H(t)=H_C+H_D(t)+H_T(t)
\end{equation}
where
\begin{eqnarray}
H_C &=& \sum_{k\alpha\sigma}(\epsilon_{k\alpha}-\mu_{\alpha})c^{\dagger}_{k\alpha\sigma}c_{k\alpha\sigma} \nonumber \\
H_D(t)&=& \sum_{\sigma}[\epsilon_{dot}(t)+\lambda(a+a^{\dagger})]d^{\dagger}_{\sigma}d_{\sigma}
+Ud^{\dagger}_{\uparrow}d_{\uparrow}d^{\dagger}_{\downarrow}d_{\downarrow} \nonumber \\
& & +\omega_0 a^{\dagger}a \nonumber \\
H_T(t)&=& \sum_{k\alpha\sigma}(V_{k\alpha}(t)c^{\dagger}_{k\alpha\sigma}d_{\sigma}+h.c.).
\end{eqnarray}
The operators $d_{\sigma}^{\dagger}(d_{\sigma})$ 
and $c_{k\alpha\sigma}^{\dagger}(c_{k\alpha\sigma})$ with 
$\alpha$=L,R create(annihilate) an electron of spin $\sigma$ 
within the dot and in the left(L) and right(R) contacts 
respectively. $V_{k\alpha}$ and $\mu_{\alpha}$ represent 
the hopping amplitudes and chemical potentials whereas
$a^{\dag}(a)$ creates(annihilates) a phonon.
$\lambda$ is the strength of the electron-phonon 
interaction and $\omega_0$ is the phonon frequency.
Throughout this letter, we will be using 
atomic units where $\hbar=k_B=e=1$.
Under the assumption that the hopping matrix elements
are equal and have no explicit time and energy
dependence, coupling of the dot to the contacts 
can be parameterized as $\Gamma(\epsilon)=\bar{\Gamma}\rho(\epsilon)$ 
where $\bar{\Gamma}$ is a constant defined by 
$\bar{\Gamma}=2\pi|V(\epsilon_f)|^2$ and
$\rho(\epsilon)$ is the density of states
of the contacts. In the following, we will take
parabolic density of states and same bandwidth 
for both contacts.

We will be interested in strong coupling regime.
This amounts to resonant tunneling which entails
longer electron lifetime and strong electron-phonon
interaction. Tunneling electrons create and destroy
a phonon cloud and this leads to polaron formation
at the junction. Phonon mode is not coupled to the
leads because it is unrealistic to allow polaron
formation in the leads which are made of metal.
This model has long been used to study inelastic 
electron transport in molecular junctions in steady
state \cite{GalperinetAl07JPCM}.

Upon applying the unitary Lang-Firsov canonical
transformation in order to eliminate the electron-phonon
coupling term when the electron-phonon coupling is
sufficiently strong compared to the tunnel couplings
\cite{Goker11JPCM}, the dot Hamiltonian becomes
\begin{eqnarray}
\overline{H}_D(t) &=& S H_D(t) S^{\dagger} \nonumber \\
&=&\sum_{\sigma} \overline{\epsilon}_{dot}(t) d^{\dagger}_{\sigma}d_{\sigma} \nonumber \\
& &+\overline{U} d^{\dagger}_{\uparrow}d_{\uparrow}d^{\dagger}_{\downarrow}d_{\downarrow}
+\omega_0 a^{\dagger}a.
\end{eqnarray}
In this Hamiltonian, the dot level and the 
Hubbard interaction strengths are renormalized as 
$\overline{\epsilon}_{dot}(t)=\epsilon_{dot}(t)-(\lambda^2/\omega_0)$
and $\overline{U}=U-(2\lambda^2/\omega_0)$ respectively 
due to the canonical transformation. When we apply the 
slave boson transformation to this Hamiltonian in 
$\overline{U} \rightarrow \infty$ limit, double occupancy
of the dot level is prevented. However, standard diagrammatic
techniques are not applicable anymore. We can overcome this
problem by rewriting the original electron operator on the dot 
in terms of slave boson and pseudofermion operators as 
\begin{eqnarray}
d_{\sigma}(t)&=& b^{\dagger}(t)f_{\sigma}(t) \nonumber \\
d^{\dagger}_{\sigma}(t) &=& f^{\dagger}_{\sigma}(t)b(t)
\end{eqnarray}
subject to the restriction
\begin{equation}
Q=b^{\dagger}b+\sum_{\sigma}f^{\dagger}_{\sigma}f_{\sigma}=1,
\end{equation}
which guarantees the single occupancy of the dot level. 
The slave boson Hamiltonian turns out to be 
\begin{eqnarray}
\overline{H}(t)&=& \sum_{k\alpha\sigma}(\epsilon_{k\alpha}-\mu_{\alpha})c^{\dagger}_{k\alpha\sigma}c_{k\alpha\sigma}
+ \sum_{\sigma} \overline{\epsilon}_{dot}(t)f^{\dagger}_{\sigma}f_{\sigma}+\omega_0 a^{\dagger}a \nonumber \\
& & +\sum_{k\alpha\sigma}(\tilde{V}_{k\alpha}(t)c^{\dagger}_{k\alpha\sigma}f_{\sigma}b^{\dagger}+h.c.),
\end{eqnarray}
where the tunnel coupling is renormalized and given by
\begin{equation}
\tilde{V}_{k\alpha}(t)=V_{k\alpha}(t)exp \left[-\frac{\lambda^2}{\omega_0^2}\left(N_{ph.}+\frac{1}{2}\right)\right].
\end{equation}

Using the slave boson and pseudofermion decomposition 
of the original fermion operators on the dot, 
the retarded Green function can be expressed as \cite{Goker11JPCM}
\begin{eqnarray}
G^R(t,t_1) &=& -i\theta(t-t_1)[G^R_{pseudo}(t,t_1)B^<(t_1,t)e^{\phi(t_1-t)} \nonumber \\
& & +G^<_{pseudo}(t,t_1)B^{R}(t_1,t)e^{\phi(t-t_1)}]
\end{eqnarray}
In this expression, the phase factor is defined by
\begin{eqnarray}
\phi(t_1-t)&=& -g[N_{ph.}(1-e^{-i\omega_0 (t_1-t)})+ \nonumber \\
& &(N_{ph.}+1)(1-e^{i\omega_0(t_1-t)})],
\end{eqnarray}
where $g$ is defined as  $g=\frac{\lambda^2}{\omega_0^2}$ 
and $N_{ph.}$, given by Bose-Einstein distribution 
$N_{ph.}=\frac{1}{e^{\frac{\hbar\omega_0}{k_B T}}-1}$ 
function, represents the average number of phonons for 
temperature $T$ and phonon frequency $\omega_0$.

\begin{figure}[htb]
\centerline{\includegraphics[angle=0,width=9.2cm,height=6.6cm]{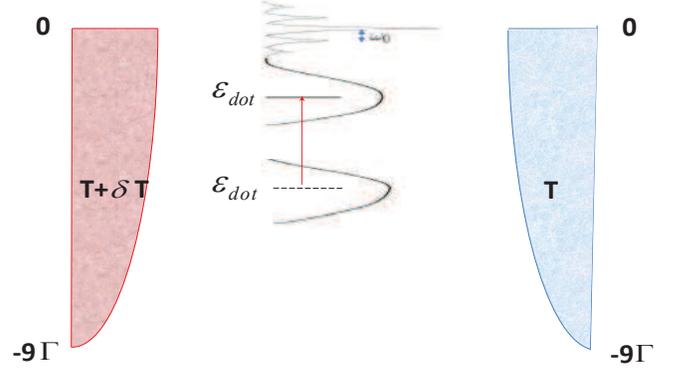}}
\caption{
This figure shows the density of states of both contacts and
the quantum dot in the initial and final states schematically. 
The temperature gradient between the contacts is also depicted. 
}
\label{Schematic}
\end{figure}

Double time retarded and lesser Green's
functions for pseudofermions and slave bosons 
are determined by solving coupled Dyson
equations in a cartesian two-dimensional grid.
We note that the phonon phase factors remain 
attached to the pseudofermion Green's functions 
in Dyson equations so that the electron-phonon
interaction is accounted for properly \cite{WerneretAl07PRL}.
The only remaining ingredient to obtain a
closed set of equations is the self energies.
We resort to non-crossing approximation(NCA) to 
express the pseudofermion and slave boson 
self-energies \cite{ShaoetAl194PRB,IzmaylovetAl06JPCM}. 
NCA is known to provide accurate 
results for dynamical quantities except
temperatures below $T/T_K\approx$0.1 or 
finite magnetic fields. We will avoid these 
regimes here. The values of the double time 
Green functions are kept in a square matrix 
which is propagated diagonally to describe 
the time evolution in an accurate way. 

In linear response, conductance of the device is given by
\begin{equation}
G(t)=\frac{L_{11}}{T}
\end{equation}
and the thermopower can be expressed as
\begin{equation}
S(t)=\frac{L_{12}(t)}{T L_{11} (t)},
\label{Seebeck}
\end{equation}
where the Onsager relations are
\begin{eqnarray}
& & L_{11} (t)= T \times \nonumber \\
& & Im \left(\int_{-\infty} ^t dt_1 \int \frac{d\epsilon}{2\pi} e^{i\epsilon(t-t_1)} \Gamma(\epsilon) G^r (t,t_1) \frac{\partial f(\epsilon)} {\partial \epsilon}\right)
\label{Onsager1}
\end{eqnarray}
and
\begin{eqnarray}
& &L_{12} (t)= T^2 \times \nonumber \\
& & Im \left(\int_{-\infty} ^t dt_1 \int \frac{d\epsilon}{2\pi} e^{i\epsilon(t-t_1)} \Gamma(\epsilon) G^r (t,t_1) \frac{\partial f(\epsilon)} {\partial T}\right).
\label{Onsager2}
\end{eqnarray}

Eq.~(\ref{Seebeck}) alongside with Eq.~(\ref{Onsager1})
and Eq.~(\ref{Onsager2}) is the central result of this work
and we will explore its consequences in the following.
We previously studied the behaviour of instantaneous 
conductance $G(t)$ for a vibrating quantum dot suddenly 
shifted into the Kondo regime in detail \cite{Goker11JPCM}. 
It is our intention in this paper to extend this 
analysis and shed light on the instantaneous thermopower
$S(t)$ for the same system. However, we will need to
stay in linear response since the Onsager relations are 
valid only in this regime.

Kondo effect is a leitmotif of many body physics
occurring as a result of hybridization between the
net spin localized inside the quantum dot and the 
continuum electrons in the leads. When the dot level 
is situated below the Fermi level at low temperatures, Kondo
resonance emerges as a very sharp resonance pinned 
to the Fermi level of the contacts. A low energy scale 
called Kondo temperature provides a good estimate for 
the linewidth of the Kondo resonance. Kondo temperature
is denoted with $T_K$ and given by
\begin{equation}
T_K \approx \left(\frac{D\Gamma}{4}\right)^\frac{1}{2}
\exp\left(-\frac{\pi|\overline{\epsilon}_{\rm dot}|}{\Gamma}\right),
\label{tkondo}
\end{equation}

In Eq.~(\ref{tkondo}) $D$ is the half bandwidth of the 
conduction electrons while $\Gamma=\bar{\Gamma} \rho(\epsilon_f)$.

\begin{figure}[htb]
\centerline{\includegraphics[angle=0,width=8.0cm,height=5.5cm]{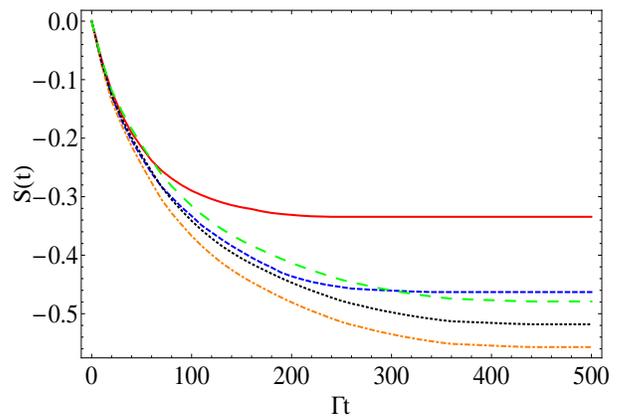}}
\caption{
This figure shows the instantaneous thermopower $S(t)$ 
immediately after the dot level has been moved to its 
final position for $T_1$=0.0035$\Gamma$ (red solid), 
$T_2$=0.0028$\Gamma$ (blue short dashed),
$T_3$=0.0021$\Gamma$ (orange dot dashed), 
$T_4$=0.0014$\Gamma$ (black dotted), 
$T_5$=0.0007$\Gamma$ (green long dashed)
in linear response without any electron-phonon coupling.
}
\label{Fig2}
\end{figure}

We will consider the behaviour of the instantaneous 
value of the thermopower immediately after the dot 
level is switched from $\epsilon_1=-5\Gamma$ to 
$\epsilon_2=-2\Gamma$ at t=0 via a gate voltage. 
A transition is triggered from a non-Kondo state 
to a Kondo state as a result of this abrupt movement. 
The density of states of the dot both in initial and 
final levels as well as the parabolic structure of
the density of states of the contacts is shown 
schematically in Fig.\ref{Schematic}. Since it takes 
considerable amount of time for the Kondo resonance 
to form in the final state \cite{NordlanderetAl99PRL}, 
dynamical quantities would require a similar amount 
of time to adjust and reach their steady state values. 
Consequently, their evolution in the long timescale 
is non-trivial and requires a careful analysis.

Instantaneous thermopower is displayed for various ambient 
temperatures in Fig. \ref{Fig2} after switching to the final
dot level in infinitesimal bias for zero electron-phonon
coupling. In this figure, we clearly observe
that the thermopower starts decaying from zero to its 
steady state value for all temperatures studied. After
a certain decay time, it reaches its steady state value.
This steady state is negative for all temperature
values. However, its absolute value increases until a
certain temperature. Once we increase the temperature
beyond that, the steady state value starts decreasing but
the decay time stays constant. The temperature at which
the steady state thermopower reaches its minimum
value has been previously identified as the Kondo
temperature \cite{CostietAl10PRB} and our results 
are in agreement with this observation. The extra 
insight our results provide is the saturation
of the decay time of thermopower and inverse of
the saturated decay time. This non-trivial
and counter-intuitive result is shown quite strikingly
in Fig. \ref{Fig4}. We will elaborate on its microscopic 
nature later on. Both the temperature below which 
saturation occurs and the inverse of the saturated 
decay time are equal to the Kondo temperature.
These novel observations can become a powerful 
tool to determine the Kondo temperature during
experiments involving thermoelectric switching 
of a single electron transistor. 

\begin{figure}[htb]
\centerline{\includegraphics[angle=0,width=8.2cm,height=5.6cm]{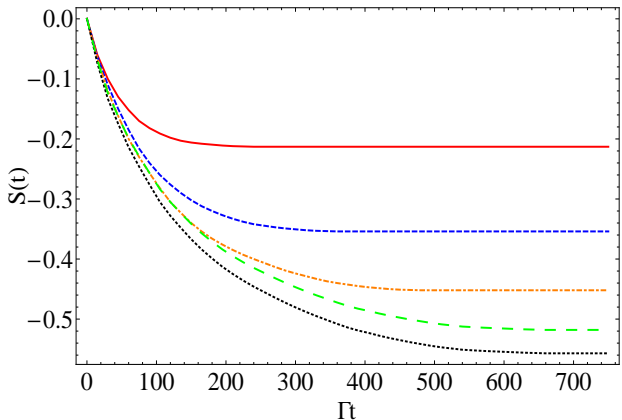}}
\caption{
This figure shows the instantaneous thermopower $S(t)$ 
immediately after the dot level has been moved to its 
final position for $T_1$=0.0035$\Gamma$ (red solid),
$T_2$=0.0028$\Gamma$ (blue short dashed),
$T_3$=0.0021$\Gamma$ (orange dot dashed), 
$T_4$=0.0014$\Gamma$ (black dotted), 
$T_5$=0.0007$\Gamma$ (green long dashed)
in linear response for $g$=2.25.
}
\label{Fig3}
\end{figure}

In Fig.~\ref{Fig3}, we investigate the instantaneous 
thermopower for the same parameters used in Fig.~\ref{Fig2}, 
but this time we use $g$=2.25 with $\omega_0$=0.08$\Gamma$.
In this case, the dot level is shifted slightly downwards 
due to renormalization leading to a smaller Kondo temperature
compared to $g$=0 case. As a result, steady state thermopower
values are smaller than $g$=0 case if the ambient
temperature is above both Kondo temperatures. However, 
they are larger than $g$=0 case when the ambient 
temperature is lowered below both Kondo temperatures. 

Kondo temperature can once again be identified as 
the temperature at which steady state thermopower 
reaches its minimum value. This is also 
the same temperature below which decay time 
saturates in analogy with $g$=0 case. The major 
difference with $g$=0 case lies in the fact that the 
saturated decay time is now much longer as one can 
clearly see for the lowest curves in both Fig.~\ref{Fig2} 
and Fig.~\ref{Fig3} because the Kondo temperature
is much smaller as a result of renormalization of
the dot level. In fact, our calculations 
show that the decay time of thermopower scales with 
$1/T_K$. That is why curves with and 
without electron-phonon coupling perfectly 
overlap in Fig.~\ref{Fig4}. This scaling has been 
previously observed for time dependent conductance 
calculations in the absence of any electron-phonon 
coupling as well \cite{PlihaletAl05PRB}. Same underlying 
mechanism holds for thermopower as well because both 
dynamical quantities depend on the evolution of Kondo 
resonance whose development has been shown to be inversely 
proportional to the Kondo temperature \cite{NordlanderetAl99PRL}.

\begin{figure}[htb]
\centerline{\includegraphics[angle=0,width=8.3cm,height=5.7cm]{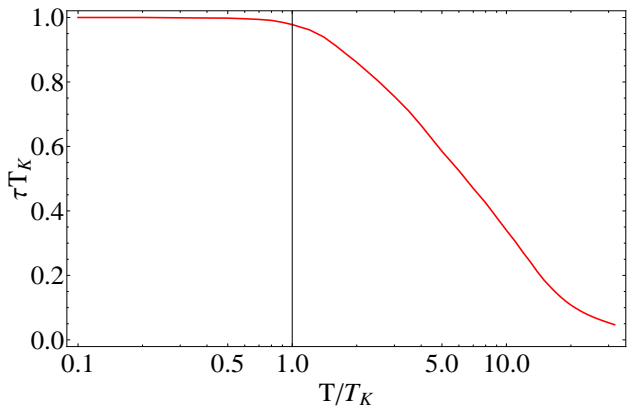}}
\caption{
This figure shows the decay time of
thermopower $\tau$ as a function of
temperature for both zero and finite
electron-phonon coupling strengths.
Saturation of the decay time of thermopower
around $T_K$ is clearly visible here. 
}
\label{Fig4}
\end{figure}

Electron-phonon coupling strength $\lambda$ is generally
unknown in an experiment. These type of measurements may
pave the way for determination of $\lambda$ if
the phonon frequency $\omega_0$ is known. Once the
Kondo temperature is identified via the saturation
of decay time described above, Eq.~(\ref{tkondo})
can be solved for renormalized dot level 
$\overline{\epsilon}$. This would in turn 
enable to find $\lambda$ because the final bare
dot level is adjusted via gate voltage thus known.

It is possible to understand previous numerical results 
intuitively with the help of Sommerfeld expansion
at low temperatures. It is given by
\begin{equation}
S(T)=-\frac{\pi^2 T}{3 A(0,T)}\frac{\partial A}{\partial \epsilon} \rvert_{\epsilon=0}
\end{equation}
in atomic units where $A(0,T)$ is the value of
the spectral function at Fermi level and
$\frac{\partial A}{\partial \epsilon}$ is its 
derivative. Since the Kondo resonance lies
slightly above the Fermi level \cite{CostietAl94JPCM},
the derivative of the dot density of states at
Fermi level is positive. This results in 
negative thermopower values for all temperatures
in Kondo regime due to the negative sign at the 
beginning of Sommerfeld expansion.

When T$\le T_K$, the Kondo resonance is fully formed 
in steady state and lowering the temperature any further
doesn't alter its final shape. Similarly, evolution to 
that final shape takes same amount of time for any $T$ 
below $T_K$. Hence, the decay time of thermopower to its
steady state stays constant. However, the steady
state value keeps going down in magnitude because of 
$T$ prefactor in Sommerfeld expansion. This prefactor 
becomes the only variable below $T_K$ in Sommerfeld expansion 
governing the steady state value since the final shape
of spectral function is the same. Subsequently, steady
state thermopower curve is nonmonotonic and goes to zero
at $T$=0 eventually. This issue has been confirmed 
by other sources too \cite{CostietAl10PRB}. On the 
other hand, the Kondo resonance is simply nonexistent 
when the dot is in its initial level since T$\gg T_K$. 
Consequently, dot density of states is essentially flat 
at Fermi level giving rise to zero slope which implies
zero thermopower. This is why $S(t)$=0 at $t$=0 and
this is in good agreement with previous steady
state results \cite{CostietAl10PRB}.

In conclusion, we investigated the temporal evolution
of the thermopower in response to an abrupt movement
of the dot level to a position where the Kondo effect 
is present. We identified both the temperature below
which decay time saturates and the inverse of the 
saturated decay time as new tools to identify the 
value of the Kondo temperature and the electron-phonon 
coupling strength. Transient dynamics of thermopower
offers a complementary picture to the previous
steady state results because providing a detailed analysis
of turning this device on and off is crucial for
practical device characterization. 

In principle, it should be possible to reproduce
short time behaviour of our results with time dependent
density matrix renormalization group method, however we
do not believe it would be sufficient to capture the long
timescale governing the development of Kondo resonance 
since the number of states required to keep the truncation
error fixed grows exponentially with time making calculations
involving long times unfeasible \cite{FeiguinetAl105PRB}.

We believe that it is possible to perform this 
experiment with present day technology because 
latest advances in ultrafast pump-probe techniques 
enable to measure the transients down to femtosecond
timescale \cite{Teradaetal10JPCM,TeradaetAl10Nature}.
The entire set-up shown in Fig.\ref{Schematic} is kept
in a dilution refrigerator in order to access the 
ambient temperatures around $T_K$. This means that
both electrodes are in principle at same ambient 
temperature. One needs to induce a temperature gradient
between the contacts in this experiment by irradiating
one of the contacts with a laser beam. Consequently, we 
aim to motivate further experiments in this field with 
this paper.

AG would like to acknowledge several fruitful
discussions with Dr. Xuhui Wang during the initial stages
of this project and both authors thank T$\ddot{u}$bitak for
generous financial support via grant 111T303.

\bibliographystyle{iopams}

\providecommand{\newblock}{}

\end{document}